\documentclass[journal,10pt,twocolumn]{IEEEtran}
\usepackage{xmpaper,lzqsymbol}
\usepackage{amssymb,amsmath,latexsym}
\usepackage{graphicx,subfigure}
\usepackage{color,cite,times}
\usepackage{epstopdf}
\usepackage{multirow}
\usepackage{array}

\usepackage{amsmath}
\usepackage{bm}
\usepackage{algorithm}
\usepackage{algorithmic}
\usepackage{bm}
\usepackage[T1]{fontenc}

\usepackage{booktabs}
\usepackage{bbm} 
\usepackage{subfigure}

\def\baselinestretch{0.945}
\begin{document}
\title{Hybrid Offline-Online Design for Reconfigurable Intelligent Surface Aided UAV Communication}
\DeclareRobustCommand*{\IEEEauthorrefmark}[1]{%
    \raisebox{0pt}[0pt][0pt]{\textsuperscript{\footnotesize\ensuremath{#1}}}}

\author{Kaiyuan Tian, Bin Duo,~\IEEEmembership{Member, IEEE}, Xiaojun~Yuan,~\IEEEmembership{Senior~Member, IEEE}, and Wu Luo,~\IEEEmembership{Member, IEEE}.
\thanks{
K.~Tian and W.~Luo are with the  State Key Laboratory of Advanced Optical Communication Systems and Networks, Department of Electronics, Peking University.(e-mail: tiankaiyuan@stu.pku.edu.cn; luow@pku.edu.cn).
B.~Duo, and X.~Yuan are with the National Laboratory of Science and Technology on Communications, University of Electronic Science and Technology of China, Chengdu 611731, China (e-mail: duobin@cdut.edu.cn; xjyuan@uestc.edu.cn).
B.~Duo is also with the College of Information Science and Technology, Chengdu University of Technology, Chengdu 610059, China.
}
}

\maketitle

\begin{abstract}
 This letter considers the reconfigurable intelligent surface (RIS)-aided unmanned aerial vehicle (UAV) communication systems in urban areas under the general Rician fading channel.  A hybrid offline-online design is proposed to improve the system performance  by leveraging both the statistical channel state information (S-CSI)  and instantaneous channel state information (I-CSI).  For the offline phase, we aim to maximize the expected average achievable rate based on the S-CSI by jointly optimizing the RIS's phase-shift and UAV trajectory. The formulated stochastic optimization problem is difficult to solve due to its  non-convexity. To tackle this problem, we propose an efficient algorithm by leveraging the stochastic successive convex approximation (SSCA) techniques. For the online phase, the UAV adaptively adjusts the transmit beamforming and user scheduling according to the effective I-CSI. Numerical results verify that the proposed hybrid design  performs better than various bechmark schemes, and also demonstrate a favorable trade-off between system performance and CSI overhead.

\end{abstract}

   \section{Introduction}
 For the sixth-generation (6G) wireless communications,  the new concept of reconfigurable radio environment was proposed to overcome the random and time-varying channels, which can be achieved by two emerging techniques: unmanned aerial vehicle (UAV) and reconfigurable intelligent surface (RIS) \cite{you2021}. Specifically, UAVs operating at low altitudes are with high flexibility and maneuverability, and its trajectory can be designed to create favorable communication channels with ground users. Besides, RIS is a programmable metamaterial  surface consisting of many low-cost passive reflecting elements, changing the propagation direction of electromagnetic waves by adjusting the phase shifts of elements\cite{guo2020we}.

   Recently, the RIS technique has been integrated into UAV communication systems to reconfigure the propagation environments of air-to-ground channels, thus extending the communication service coverage and enhancing the communication quality \cite{li2020re, K2022hy, Ranjha2021, Y2022fa}. In \cite{li2020re}, the authors studied the maximum average achievable rate problem in RIS-aided UAV communication systems by jointly optimizing the UAV trajectory and RIS's phase-shift. In \cite{K2022hy}, the RIS was also applied to aid the full-duplex UAV communication systems that considered both uplink and downlink transmissions. The authors in \cite{Ranjha2021} studied the ultrahigh reliability problem for internet of things devices communicating by both UAV relay and RIS. Besides, the authors in \cite{Y2022fa} considered the fair downlink communications for RIS-UAV enabled mobile vehicles.

  It is worth pointing out that the aforementioned studies adopted the deterministic line-of-sight (LoS)-dominant channel model, which is usually a valid assumption for rural areas. As a result, these works considered the deterministic optimization problems and proposed algorithms working in an offline manner. However, empirical studies showed that the Rician factor, which is the ratio of the power of the deterministic LoS component to the stochastic NLoS component, may drop to  $-5~\rm{dB}$  in the low altitude platform of urban areas \cite{X2020an}. In these scenarios, the existing offline design may suffer considerable rate performance loss, since it cannot adapt to the random and dynamic UAV-ground channel states\cite{C2020hy}. Although some recent works have adopted the Rician fading channel model  in RIS-aided UAV communication systems (e.g., \cite{li2021ro, hua2021, Z2021su }), they still followed the offline design based on the deterministic optimization, which is not suitable for the environments with relatively small Rician factor. Note that  the deep Q-network (DQN)  based online design was utilized in RIS-aided UAV communication systems, such as in \cite{X2021ma}. However, these works mainly focused on the  obstacle avoidance and random user movement instead of the uncertain channel states. Besides,  they need to obtain the full instantaneous channel state information (I-CSI) between UAV, RIS  and users in real-time, which is quite challenging due to the mobility of the UAV and the passivity of the RIS \cite{Z2021su}. Thus, it is important to investigate how to improve the system performance for RIS-aided UAV communication systems under the general Rician fading channel model with limited CSI overhead.





Motivated by the above, we propose an efficient algorithm for the multi-user RIS-aided UAV communication systems in both offline and online phases,\footnote{Note that although the authors in \cite{C2020hy} and \cite{C2020en} also utilized a combined
offline and online design approach,  the system model and the proposed algorithm are different when integrating RIS into UAV communication systems.} which is easier to implement in the urban areas. Specifically, in the offline phase, our goal is to maximize the \emph{expected average achievable rate} from all the users according to the statistical CSI (S-CSI). To tackle this problem, we propose an  iterative algorithm based on the stochastic successive convex approximation (SSCA) techniques \cite{A2018on}. For the online phase, with the fixed RIS's phase-shift and UAV trajectory, the transmit beamforming and user scheduling are dynamically designed to  accommodate the effective I-CSI,  and the computational complexity of the online design facilitates the practical implementation. Our simulation results demonstrate that the proposed hybrid design achieves a better system performance with limited CSI overhead.


    \section{System Model }
    We consider a RIS-aided UAV communication downlink transmission system consisting of a rotary-wing UAV equipped with a uniform linear array (ULA) with $N_t$ elements , $K$ single-antenna ground users, and a RIS mounted on a building.

      The UAV acts as an aerial base station (BS) and flies at a fixed altitude $z_F$ within a given period $T$ to meet the communication service requirements of users. To facilitate the  design of the UAV trajectory, \emph{T} is equally divided into \emph{N}  time slots with step size $\delta_t$, i.e., $T = N\delta_t$. Therefore, the UAV trajectory can be approximated by the sequence $\mathbf{q}[n]=[x[n], y[n]]^T, n \in \mathcal{N}=\{1,\cdots, N \}$. With a given maximum UAV speed $v_{\max}$, $N$ can be chosen properly such that the time for UAV location changing within $\delta_t$ can be negligible. Accordingly, the UAV trajectory should satisfy the constraints as
  \begin{small}  \begin{subequations} \label{mobility.1}
    \begin{align}
        & ||\mathbf{q}[n]-\mathbf{q}[n-1]||^{2} \leq D^2, \forall n \label{mobility.1.a}\\
        & \mathbf{q}[0] = \mathbf{q}_0, \mathbf{q}[N]=\mathbf{q}_F, \label{mobility.1.b}
    \end{align}
    \end{subequations} \end{small}where $D=v_{\max}\delta_t$ is the maximum horizontal distance that the UAV can move within a single time slot, and $\mathbf{q}_0$ and $\mathbf{q}_F$ denote the UAV's initial and final horizontal locations, respectively.

   The RIS is deployed to enhance the system performance, which is equipped with a uniform rectangular array (URA)  containing  $M=M_x\times M_z$ reflecting elements. The RIS's phase-shift in the $n$th time slot is defined by a diagonal matrix $\boldsymbol{\Theta}[n]=\diag\{ \theta_1[n],  \theta_2[n],\cdots,\theta_M[n] \}$, where $\theta_m[n]=e^{j\varphi_m[n]}$  and $\varphi_m[n] \in \left[0,2\pi\right), m \in \mathcal{M}=\{1,\cdots, M\}$, is the phase shift of the $m$th reflecting element  in the $n$th time slot.

Following \cite{li2021ro, Z2021su}, we assume the Rician fading channel model for all communication links. Specifically, the channel of the RIS-ground user links can be modeled as
\begin{footnotesize}
\begin{align}
\mathbf{h}_{r, k}[n]=\sqrt{\rho d_{R\! G,k}^{-\alpha}[n]}\left(\sqrt{\frac{\beta_{R\! G}}{1+\beta_{R \!G}}} \overline{\mathbf{z}}_{r, k}[n]+\sqrt{\frac{1}{1+\beta_{R\! G}}} \mathbf{z}_{r, k}[n]\right),
\end{align}\end{footnotesize}where  $d_{R\! G,k}[n]$ is the distance between the RIS and ground user $k$ in the $n$th time slot, $\rho$ is the path loss at the reference distance $D_0=1m$, and $\mathbf{z}_{r, k}[n] \in \mathbb{C}^{M \times 1}$ has  circularly symmetric complex Gaussian (CSCG) entries with zero mean and unit variance accounting for small-scale fading in the $n$th time slot. Moreover, $\alpha$, $\beta_{R\!G}$ and $\overline{\mathbf{z}}_{r, k}[n]$ denote the path loss exponent,  Rician factor and  deterministic component of the RIS-ground user links in the $n$th time slot, respectively.

The UAV-RIS and UAV-ground user links in the $n$th time slot can be generated with a similar procedure i.e.,
\begin{footnotesize}\begin{equation}
\mathbf{G}[n] =\sqrt{\rho d_{U\! R}^{-\gamma}[n]}\left(\sqrt{\frac{\beta_{U\! R}}{1+\beta_{U \!R}}} \overline{\mathbf{Z}}[n]+\sqrt{\frac{1}{1+\beta_{U\! R}}}  \mathbf{Z}[n]\right),
\end{equation}\end{footnotesize}
\begin{footnotesize}\begin{equation}
\mathbf{h}_{d, k}[n] =\sqrt{\rho d_{U \!G,k}^{-\kappa}[n]}\left(\sqrt{\frac{\beta_{U \!G}}{1+\beta_{U \!G}} \overline{\mathbf{z}}_{d, k}}[n]+\sqrt{\frac{1}{1+\beta_{U \!G}}}  \mathbf{z}_{d, k}[n]\right),
\end{equation}\end{footnotesize}where $\overline{\mathbf{Z}}[n]$ and $\overline{\mathbf{z}}_{d, k}[n]$ denote the deterministic components  in the $n$th time slot, and $\mathbf{Z}[n] \in \mathbb{C}^{M \times N_t}$ and $\mathbf{z}_{d, k}[n] \in \mathbb{C}^{N_t \times 1}$ are the corresponding small-scale fading components  in the $n$th time slot similarly to $\mathbf{z}_{r, k}[n]$.

We adopt the  time-division multiple access (TDMA) manner for the users, which means that the UAV can only communicate with only one user at one time slot. Define a binary variable $a_k[n]$, which indicates that user $k$ is served by the UAV in the $n$th time slot if $a_k[n]= 1$, and $a_k[n]= 0$, otherwise. Then, we have the following scheduling constraints
   \begin{small}  \begin{subequations} \label{tdma}
    \begin{align}
&\sum_{k=1}^{K} a_{k}[n] \leq 1, \forall k, n \label{tdma.1.a}\\
&a_{k}[n] \in\{0,1\}. \forall k, n \label{tdma.1.b}
    \end{align}
    \end{subequations}\end{small}If the UAV communicates with user $k$ in the $n$th time slot, the achievable rate (bits/second/Hertz) can be expressed as
     \begin{small}
    \begin{equation} \label{rate.111}
     R_k[n]=\log _{2}\left(1+\frac{\left|\left(\mathbf{h}_{r, k}^{H}[n] \mathbf{\Theta}[n] \mathbf{G}[n]+\mathbf{h}_{d, k}^{H}[n]\right)\mathbf{w}[n]\right|^{2}}{\sigma^{2}}\right),
    \end{equation} \end{small}where $\mathbf{w}[n]\in \mathbb{C}^{N_t \times 1}$ is the transmit beamforming of the UAV in the $n$th time slot, and $\sigma^2$ is the noise variance.
\section{Problem Formulation}
Our objective is to maximize the average achievable rate via jointly optimizing the transmit beamforming $\mathbf{W} \triangleq \{\mathbf{w}[n], n\in\mathcal{N}\}$, the RIS's phase-shift $\boldsymbol{\Phi}\triangleq \{\boldsymbol{\Theta}[n], n\in\mathcal{N}\}$, the user scheduling $\mathbf{A}\triangleq \{a_{k}[n], n\in\mathcal{N}, k\in\mathcal{K}\}$, and the UAV trajectory $\mathbf{Q}\triangleq \{\mathbf{q}[n], n\in\mathcal{N}\}$ over the entire $N$ time slots. Thus, the optimization problem can be expressed as
    \begin{small}  \begin{subequations}\label{optimal.111}
    \begin{align}
          (\mathrm{P_A})&  \max\limits_{\mathbf{Q}, \mathbf{A}, \boldsymbol{\Phi}, \mathbf{W}} \quad
         \frac {1}{N}\sum \limits_{k=1}^K  \sum \limits_{n=1}^N  a_{k}[n] R_k[n] \\
         & ~~\textrm{s.t.} \quad \quad~~~\left|\theta_m[n]\right| = 1, \forall n, m, \label{7b}\\
         & ~~~~~~~~\quad~~~\left\|\mathbf{w}[n]\right\|^{2} \leq P , \forall n,\label{7c}\\
         & ~~~~ \quad\quad ~~~~~ \eqref{mobility.1},\eqref{tdma} \nonumber,
     \end{align}
    \end{subequations} \end{small}where $P$ is the maximum transmit power of the UAV. The optimal solution to problem $(\mathrm{P_A})$ is difficult to obtain, since it is impossible to acquire the accurate I-CSI $\mathbf{F}[n] \triangleq\{\mathbf{G}[n], \mathbf{h}_{r, k}[n], \mathbf{h}_{d, k}[n]\}$ at all possible UAV locations in the region of interest before the flight. Considering this difficulty in practice, we propose a hybrid offline-online design for RIS-aided UAV communication systems, consisting of the following two phases.


\textbf{Offline phase:}  Prior to the UAV's flight, we offline design the UAV trajectory and RIS's phase-shift  to maximize the expected average achievable rate based on  the Rician fading channel model and user's location. Therefore, the resulting RIS's phase-shift and the UAV trajectory are statistically favorable solutions. The optimization problem of the offline phase is formulated as
   \begin{small} \begin{subequations}\label{optimal.1}
    \begin{align}
          (\mathrm{P_B})& \max\limits_{\mathbf{Q}, \mathbf{A}, \boldsymbol{\Phi}, \mathbf{W}} \quad
        \mathbb{E}\left\{ \frac {1}{N}\sum \limits_{k=1}^K  \sum \limits_{n=1}^N    a_{k}[n] R_k[n] \right\}\\
         & ~~~\textrm{s.t.} \quad ~~~~~\eqref{mobility.1},\eqref{tdma}, \eqref{7b},  \eqref{7c}, \nonumber
     \end{align}
    \end{subequations}\end{small}where  $\mathbb{E}$  is the expectation operator.

\textbf{Online phase:} Since the number of transmit antennas at the UAV  is usually much smaller than the elements of the RIS, the effective I-CSI $\mathbf{h}_{k}[n] \triangleq \mathbf{G}^{H}[n] \mathbf{\Theta}[n] \mathbf{h}_{r, k}[n]+\mathbf{h}_{d, k}[n]$  usually has a much smaller dimension than the full I-CSI $\mathbf{F}[n]$ \cite{M2021I}. Thus, we only estimate the effective I-CSI  in real-time along its flight by existing methods\cite{H2020A}, and the number of channel coefficients required in each time slot can be reduced from $MN_t + MK + N_tK$ to $N_tK$. Then, the transmit beamforming and the user scheduling can be adjusted online to cater to the random environments. The optimization problem of the online phase in the $n$th time slot is formulated as
\begin{small}
    \begin{subequations}\label{optimal.2}
    \begin{align}
          (\mathrm{P_C})~&  \max\limits_{ \mathbf{w}[n],  \mathbf{a}[n]} ~
         \sum \limits_{k=1}^K a_{k}[n] \log _{2}\left(1+\frac{\left|\mathbf{h}_{k}^{H}[n] \mathbf{w}[n]\right|^{2}}{\sigma^{2}}\right)\\
         & ~~~~\textrm{s.t.} \quad ~\eqref{tdma},  \eqref{7c}\nonumber,
     \end{align}
    \end{subequations} \end{small}where  $\mathbf{a}[n] \triangleq \{a_{k}[n], k\in\mathcal{K}\}$ is the user scheduling in the $n$th time slot.

\section{Proposed Hybrid Offline-Online Design}
\subsection{Proposed  Offline Design}
For the offline design, we develop an alternating optimization algorithm to tackle the non-convex problem $(\mathrm{P_B})$.  Specifically, problem $(\mathrm{P_B})$ is partitioned into four subproblems: the optimization of the transmit beamforming, the RIS's phase-shift, the user scheduling, and the UAV trajectory.
\subsubsection{Transmit Beamforming Optimization}
   It is well-known that transmit beamforming as the maximum ratio transmission (MRT) is optimal i.e.,
  \begin{small}\begin{equation}
  \mathbf{w}^{\mathrm{opt}}[n]=\sqrt{P} \frac{\left(\mathbf{h}_{r,k}^{H}[n] \boldsymbol{\Theta}[n] \mathbf{G}[n]+\mathbf{h}_{d,k}^{H}[n]\right)^{H}}{\left\|\mathbf{h}_{r,k}^{H}[n] \boldsymbol{\Theta}[n] \mathbf{G}[n]+\mathbf{h}_{d,k}^{H}[n]\right\|}.
  \end{equation}\end{small}By substituting $\mathbf{w}^{\mathrm{opt}}[n]$ to (6), the average achievable rate of the user $k$ in the $n$th time slot can be simplified as
   \begin{small} \begin{equation}R_k[n]= \log _{2}\left(1+\frac{P}{\sigma^{2}}\left\|\mathbf{h}_{r, k}^{H}[n] \boldsymbol{\Theta} [n]\mathbf{G}[n]+\mathbf{h}_{d, k}^{H}[n]\right\|^{2}\right).  \end{equation}\end{small}
\subsubsection{RIS's Phase-Shift Optimization}
To make the optimazation of $\boldsymbol{\Phi}$ more tractable, we further define $\bm{\theta}_k[n]=[\theta_{k,1}[n],\cdots,\theta_{k,M}[n]]^{\mathrm{T}}$  and $\boldsymbol{\varphi}_k[n]=\left[\varphi_{k,1}[n], \cdots, \varphi_{k,M}[n]\right]^{\mathrm{T}}$ for the $k$th user in the $n$th time slot. Thus, with given $\mathbf{A}$ and $\mathbf{Q}$, the optimization problem of $\boldsymbol{\Phi}$   in the $n$th time slot can be written as

\begin{footnotesize}
\begin{align}
    (\mathcal{P}_{B1})~ &  \max \limits_{\boldsymbol{\varphi}}  ~  \mathbb{E}\left\{\log \left(1+\frac{P}{\sigma^{2}}\left\|\left(e^{\jmath \boldsymbol{\varphi}_k[n]}\right)^{H} \mathbf{H}_{k}[n]+\mathbf{h}_{d, k}^{H}[n] \right\|^{2}\right)\right\},
\end{align} \end{footnotesize}where \begin{small}$\mathbf{H}_{k}[n]=\operatorname{diag}\left(\mathbf{h}_{r, k}^{H}[n]\right) \mathbf{G}[n]$ \end{small}. Problem $(\mathcal{P}_{B1})$ is difficult to solve, since the objective function is non-convex and has expectation operation. To tackle this difficulty, we apply the SSCA framework in \cite{A2018on} and optimize the problem  iteratively. The details are as follows.

  First, we approximate the gradient of $ (\mathcal{P}_{B1})$ with  $I$ channel samples in one batch. Specifically,   at the $l$th iteration,\begin{small} $\mathbf{F}_{i}[n]=\left\{\mathbf{G}_i[n], \mathbf{h}_{r, k, i}[n], \mathbf{h}_{d, k, i}[n]\right\}_{i=\left\{1, \cdots, I\right\}}$\end{small} are randomly generated according to the S-CSI. The realization of the objective function in $ (\mathcal{P}_{B1})$ is given by
\begin{footnotesize}
\begin{align}
f_k(\boldsymbol{\varphi}_{k})[n]=\frac{1}{I} \sum_{i=1}^{I} \log \left(1+\frac{P}{\sigma^{2}}\left\|\left(e^{\jmath \boldsymbol{\varphi}_{k}[n]}\right)^{H} \mathbf{H}_{k,i}[n]+\mathbf{h}_{d, k,i}^{H}[n]\right\|^{2}\right).
\end{align}
\end{footnotesize}According to the chain rule, the gradient in one realization is given by
\begin{small} \begin{align}
\nabla R_{k,i}(\boldsymbol{\varphi}_{k})[n]=\operatorname{Re}\left\{-\jmath \boldsymbol{\theta}^{*}_{k}[n] \circ \nabla R_{k,i}(\boldsymbol{\theta}_k)[n]\right\},
\end{align}\end{small}where
\begin{small}$$\nabla R_{k,i}(\boldsymbol{\theta}_k)[n]=\frac{\frac{2 P}{\sigma^{2}} \mathbf{H}_{k,i}[n] \left(\mathbf{H}_{k,i}^{H}[n] \boldsymbol{\theta}_{k}[n]+\mathbf{h}_{d, k,i}^{H}[n]\right)}{1+\frac{P}{\sigma^{2}} \left\|\boldsymbol{\theta}^{H}_{k}[n] \mathbf{H}_{k,i}[n]+\mathbf{h}_{d, k,i}^{H}[n]\right\|^{2}},
$$\end{small}$\nabla f(\mathbf{x})$ is the gradient vector of function $f(\mathbf{x})$ with respect to vector $\mathbf{x}$, $ \boldsymbol{\theta}^{*}_{k}[n]$ is the conjugate  of $\boldsymbol{\theta}_{k}[n]$,  $\circ$ denotes the  Hadamard product, and $\operatorname{Re}\{\cdot\}$ is the real part of a complex number. Thus, the stochastic gradient  of problem $(\mathcal{P}_{B1})$ at the $l$th iteration is updated by
\begin{small} \begin{align}
\mathbf{f}_k[n]=\left(1-\zeta\right) {\mathbf{f}}^{(l)}_k[n]-\zeta \nabla f_k\left({\boldsymbol{\varphi}}^{(l)}_k\right)[n],
\end{align}\end{small}where \begin{small}$\nabla {f}_k\left({\boldsymbol{\varphi}}^{(l)}_k\right)[n]=-\frac{1}{I} \sum_{i=1}^{I} \nabla R_{k,i}\left({\boldsymbol{\varphi}}^{(l)}_k\right)[n]$ \end{small}is the stochastic gradient in one batch, and $\zeta=l^{-\nu}, 0.5 \leq \nu \leq 1$.

Second, we  update the surrogate function  \begin{small}${f}_{k}\left(\boldsymbol{\varphi}_k, {\boldsymbol{\varphi}}^{(l)}_k\right)[n],$ \end{small} which is constructed by the second order Taylor expansion and can be viewed as a concave approximation of ${f}_{k}(\boldsymbol{\varphi}_{k})[n]$ as


\begin{footnotesize}\begin{align}
{f}_{k}\left(\boldsymbol{\varphi}_k, {\boldsymbol{\varphi}}^{(l)}_k\right)[n] = \left\langle\boldsymbol{\varphi}_{k}[n]-{\boldsymbol{\varphi}}^{(l)}_{k}[n], \mathbf{f}_{k}[n]\right\rangle+\frac{\tau}{2}\left\|\boldsymbol{\varphi}_{k}[n]-{\boldsymbol{\varphi}}^{(l)}_{k}[n]\right\|^{2},
\end{align}\end{footnotesize}where $\langle \boldsymbol{\varphi}_{k}[n]-{\boldsymbol{\varphi}}^{(l)}_{k}[n], \mathbf{f}_{k}[n] \rangle$ denotes the inner product operation,  and $\tau>0$ is a constant representing the learning rate. Therefore, we have the following convex optimization problem
\begin{small} \begin{align}
\hat{\boldsymbol{\varphi}}_{k}[n]=\arg \min _{\forall \boldsymbol{\varphi}_{k}[n]} {f}_{k}\left(\boldsymbol{\varphi}_{k}, {\boldsymbol{\varphi}}^{(l)}_{k}\right)[n].  \label{17}
\end{align} \end{small}and the closed-form optimal solution of \eqref{17} is
\begin{small} \begin{align}
\hat{\boldsymbol{\varphi}}_{k}[n]={\boldsymbol{\varphi}}^{(l)}_{k}[n]-\frac{\mathbf{f}_{k}[n]}{\tau}.
\end{align} \end{small}

Finally, $\boldsymbol{\varphi}_{k}[n]$ at the $l$th iteration is updated by
\begin{small} \begin{align}
\boldsymbol{\varphi}_{k}[n]=\left(1-\xi\right) {\boldsymbol{\varphi}}^{(l)}_{k}[n]+\xi \hat{\boldsymbol{\varphi}}_{k}[n],
\end{align} \end{small}where $\xi=l^{-\mu}$, and $\delta<\mu \leq 1$ to guarantee convergence  \cite{A2018on}. Thus, the optimal $\boldsymbol{\varphi}[n]$ is given by
\begin{small} \begin{align}
\boldsymbol{\varphi}[n]= \sum \limits_{k=1}^K  a_{k}[n]\boldsymbol{\varphi}_{k}[n].
\end{align} \end{small}
  \subsubsection{User Scheduling Optimization}
With given $\boldsymbol{\Phi}$ and $\mathbf{Q}$, the optimization problem of  $\mathbf{A}$  can be expressed as
    \begin{small}   \begin{subequations}
    \begin{align}
          (\mathcal{P}_{B2})~ & \max\limits_{ \mathbf{A}} ~
        \mathbb{E}\left\{ \frac {1}{N}\sum \limits_{k=1}^K  \sum \limits_{n=1}^N   a_{k}[n] R_k( \mathbf{F}_{i}[n])[n] \right\}\\
         & ~~\textrm{s.t.} ~~ \eqref{tdma}\nonumber.
     \end{align}
   \end{subequations}\end{small}We use a batch of randomly generated channel samples to deal with the expectation operation in the objective function by applying  the sample average approximation method in \cite{guo2020we}. Thus, $(\mathcal{P}_{B2})$ can be approximately solved by $(\mathcal{P}_{B2.1})$.
 \begin{small}     \begin{subequations}
    \begin{align}
          (\mathcal{P}_{B2.1})~ & \max\limits_{ \mathbf{A}} ~
         \frac {1}{NI}\sum \limits_{k=1}^K  \sum \limits_{n=1}^N  \sum \limits_{i=1}^I a_{k}[n] R_k( \mathbf{F}_{i}[n])[n] \\
         & ~~\textrm{s.t.} \quad \eqref{tdma}\nonumber.
     \end{align}
    \end{subequations} \end{small}Problem $(\mathcal{P}_{B2.1})$ is a linear optimization problem by relaxing the binary variable. Thus, it can be efficiently solved by using standard optimization solvers such as CVX \cite{cvx2017}.  The details of this method can be found in \cite{hua2021}, and are omitted here for brevity.

\subsubsection{UAV Trajectory Design}
    With given $\mathbf{A}$ and $\boldsymbol{\Phi}$, the optimization problem of $\mathbf{Q}$  can be expressed as
    \begin{small}
    \begin{subequations}
    \begin{align}
          (\mathcal{P}_{B3})~& \max\limits_{\mathbf{Q}} ~
          \mathbb{E}\left\{\frac {1}{N}\sum \limits_{k=1}^K  \sum \limits_{n=1}^N   a_{k}[n] R_k(\mathbf{Q}; \mathbf{F}_{i}[n])[n] \right\}\\
         & ~~\textrm{s.t.} ~~ \eqref{mobility.1}.\nonumber
     \end{align}
    \end{subequations} \end{small}By applying the same procedure as in the user scheduling optimization, problem $(\mathcal{P}_{B3})$ can be approximately solved by problem $(\mathcal{P}_{B3.1})$.
        \begin{small}
    \begin{subequations}
    \begin{align}
          (\mathcal{P}_{B3.1})~& \max\limits_{\mathbf{Q}} ~
         \frac {1}{NI}\sum \limits_{k=1}^K  \sum \limits_{n=1}^N   \sum \limits_{i=1}^I  a_{k}[n] R_k(\mathbf{Q}; \mathbf{F}_{i}[n])[n] \\
         & ~~\textrm{s.t.} \quad \eqref{mobility.1}.\nonumber
     \end{align}
    \end{subequations} \end{small}Problem $(\mathcal{P}_{B3.1})$ is still non-convex with respect to $\mathbf{Q}$. To tackle this difficulty, we introduce slack variables $\mathbf{u}_k=\{u_k[n]\}_{n=1}^{N}$ and $\mathbf{v}=\{v[n]\}_{n=1}^{N}$, and consider  the following problem which has the same optimal solution as $(\mathcal{P}_{B3.1})$: \footnote{Since  constraints (25b) and (25c) must hold with equality at the optimal solution of problem $(\mathcal{P}_{B3.2})$, otherwise $u_k[n]$ and $v[n]$ can be increased to reduce the objective value.}
\begin{footnotesize} \begin{subequations}
\begin{align}
(\mathcal{P}_{B3.2})~ & \max _{\mathbf{Q}, \mathbf{u}_k, \mathbf{v}}  ~  \frac {1}{NI}\sum \limits_{k=1}^K  \sum \limits_{n=1}^N   \sum \limits_{i=1}^I a_{k}[n] R_k(\mathbf{Q}; \mathbf{F}_{i}[n])[n] \\
&~~\text { s.t. } ~~~~ d_{U\!  G,k}[n] \leq u_k[n], \forall k, n, \\
&~~~~~~~~~~~ d_{U\!  R}[n] \leq v[n], \forall n, \\
&~~~~~~~~~~~\eqref{mobility.1}\nonumber.
\end{align}
\end{subequations}\end{footnotesize}Within one time slot, we assume that $ \overline{\mathbf{Z}}[n]$ and $ \overline{\mathbf{z}}_{d, k}[n]$  do not change as $|x[n+1]-x[n]| \ll d_{U\! R}[n]$ and $d_{U\! G,k}[n]$, $|y[n+1]-y[n]| \ll$ $d_{U\! R}[n]$ and $d_{U\! G,k}[n]$ generally hold. Thus, $R_{k}[n]$ can be rewritten as

\begin{footnotesize} \begin{equation}
{R}_{k}[n] =  \log _{2}{\left(1+\frac{P}{\sigma^{2}}\left(\frac{A_k[n]}{(u_k[n])^{\kappa}}+\frac{B_k[n]}{(v[n])^{\gamma}}+\frac{C_k[n]}{(u_k[n])^{\frac{\kappa}{2}}(v[n])^{\frac{\gamma}{2}}}\right)\right) }
\end{equation}\end{footnotesize}where
\begin{footnotesize}
$$A_k[n]=\mathbf{h}^{(j)}_{d,k}[n](\mathbf{h}^{(j)}_{d,k}[n])^{H}(d^{(j)}_{U\!  G,k}[n])^{\kappa},$$
$$B_k[n]=\boldsymbol{\theta}^{H}_{k}[n] \mathbf{H}^{(j)}_{k}[n]  (\mathbf{H}^{(j)}_{k}[n])^{H} \boldsymbol{\theta}_{k}[n](d_{U\!  R}^{(j)}[n])^{\gamma},$$ $$C_k[n]=\left(\boldsymbol{\theta}^{H}_{k}[n] \mathbf{H}^{(j)}_{k}[n](\mathbf{h}_{d,k}^{(j)}[n])^{H}+ \mathbf{h}^{(j)}_{d,k}[n] (\mathbf{H}^{(j)}_{k}[n])^{H} \boldsymbol{\theta}_{k}[n] \right)$$$$\times(d^{(j)}_{U\!G,k}[n])^{\frac{\kappa}{2}}(d^{(j)}_{U\!  R}[n])^{\frac{\gamma}{2}}.$$
\end{footnotesize}



By taking the  first-order Taylor expansion at given points
\begin{footnotesize}$\mathbf{u}_k^{(j)}=\left\{u_k^{(j)}[n]\right\}_{n=1}^{N}, \forall k$\end{footnotesize} and \begin{footnotesize} $\mathbf{v}^{(j)}=\left\{v^{(j)}[n]\right\}_{n=1}^{N}$\end{footnotesize}, we approximate ${R}_{k}[n]$ as its lower bound. Specifically, the  first-order Taylor expansions of ${R}_{k}[n], u_k^{2}[n]$ and $v^{2}[n]$ are respectively given by
\begin{footnotesize} \begin{subequations}
\begin{align}
{R}_{k}[n]
\geq & \log _{2} D^{(j)}_k[n]+\frac{E^{(j)}_k[n]}{D^{(j)}_k[n] \ln 2}\left(u_k[n]-u^{(j)}_k[n]\right)  \nonumber \\
&+\frac{F^{(j)}_k[n]}{D^{(j)}_k[n]  \ln 2}\left(v[n]-v^{(j)}[n]\right)  \triangleq \hat{R}_{k}[n],\\
&-u^{2}_k[n] \leq (u_k^{(j)}[n])^{2}-2 u_k^{(j)}[n] u_k[n],\\
&-v^{2}[n] \leq (v^{(j)}[n])^{2}-2 v^{(j)}[n] v[n],
\end{align}
\end{subequations}\end{footnotesize}where
\begin{scriptsize}$$D^{(j)}_k[n]=1+\frac{P}{\sigma^{2}}\left(\frac{A_k[n]}{\left(u_k^{(j)}[n]\right)^{\kappa}}+\frac{B_k[n]}{\left(v^{(j)}[n]\right)^{\gamma }}+\frac{C_k[n]}{\left(u_k^{(j)}[n]\right)^{\frac{\kappa}{2}}\left(v^{(j)}[n]\right)^{\frac{\gamma}{2}}}\right), $$\end{scriptsize}\begin{scriptsize}
$$E^{(j)}_k[n]=-\frac{P}{\sigma^{2}}\left(\frac{\kappa A_k[n]}{\left(u_k^{(j)}[n]\right)^{\kappa+1}}+\frac{\frac{\kappa}{2} C_k[n]}{\left(v^{(j)}[n]\right)^{\frac{\gamma}{2}}\left(u_k^{(j)}[n]\right)^{\frac{\kappa}{2}+1}}\right),$$
$$F^{(j)}_k[n]=-\frac{P}{\sigma^{2}}\left(\frac{\gamma B_k[n]}{\left(v^{(j)}[n]\right)^{\gamma+1}}+\frac{\frac{\gamma}{2} C_k[n]}{\left(u_k^{(j)}[n]\right)^{\frac{\kappa}{2}}\left(v^{(j)}[n]\right)^{\frac{\gamma}{2}+1}}\right).$$
\end{scriptsize}As such, problem $(\mathcal{P}_{B3.2})$ can be approximated as
\begin{footnotesize}  \begin{subequations}
\begin{align}
& (\mathcal{P}_{B3.3}) \max _{\mathbf{Q}, \mathbf{u}_k, \mathbf{v}}  \frac{1}{NI} \sum \limits_{k=1}^K \sum_{n=1}^{N}  \sum_{i=1}^{I}
a_{k}[n] \hat{R}_{k}(\mathbf{Q}; \mathbf{F}_{i}[n])[n] \\
\text { s.t. } &\left(d_{U\!  G,k}[n]\right)^{2}+\left(u_k^{(j)}[n]\right)^{2}-2 u_k^{(j)}[n] u_k[n] \leq 0, \forall k, n, \\
&\left(d_{U\!  R}[n]\right)^{2}+\left(v^{(j)}[n]\right)^{2}-2 v^{(j)}[n] v[n] \leq 0, \forall n, \\
&~\eqref{mobility.1}\nonumber.
\end{align}
\end{subequations}\end{footnotesize}Problem $(\mathcal{P}_{B3.3})$ is a convex optimization problem, and thus can be solved by the CVX.

\subsection{Proposed  Online Design}
In the online phase, the transmit beamforming and user scheduling are optimized to accommodate the real-time channels by using the effective I-CSI $\mathbf{h}_{k}[n]$ in each time slot.
\subsubsection{Transmit Beamforming Optimization}
    Similar to the transmit beamforming optimization in the offline phase, when the UAV communicates with the user $k$ in the $n$th time slot, the optimized transmit beamforming can be expressed as
     \begin{footnotesize}  \begin{align}
  \mathbf{w}^{\mathrm{opt}}[n]=\sqrt{P} (\mathbf{h}_{k}^{H}[n] / \left\| \mathbf{h}_{k}^{H}[n] \right\|). \label{w}
    \end{align}\end{footnotesize}
  \subsubsection{User Scheduling  Optimization}
  With optimized $\mathbf{w}[n]$, the optimization problem of $\mathbf{a}[n]$  in the $n$th time slot can be expressed as
   \begin{footnotesize} \begin{subequations}
    \begin{align}
          (\mathcal{P}_{C1})~&  \max\limits_{ \mathbf{a}[n]}  ~
         \sum \limits_{k=1}^K a_{k}[n] \log _{2}\left(1+\frac{P}{\sigma^{2}}\left\|\mathbf{h}_{k}^{H}[n] \right\|^{2}\right)\\
         & ~~\textrm{s.t.} \quad  \eqref{tdma}\nonumber.
     \end{align}  \end{subequations}\end{footnotesize}Problem $(\mathcal{P}_{C1})$ can be optimized as the same algorithm of the user scheduling optimization in the offline phase.
\subsection{Overall Algorithm}
    The overall algorithm for solving $(\mathcal{P}_{A})$ in the offline and online phases is summarized in Algorithm 1. The computational complexity of the offline design is $\mathcal{O}(I_o((KN)+ I_s(KNM^{2})+(N^{3.5})))$  with $I_o$ and $I_s$  respectively being the iteration numbers of the overall problem and  the subproblem $(\mathrm{P}3.1)$. As for the online design, the complexity  is $\mathcal{O}(K)$ in the $n$th time slot.
       \begin{algorithm}[t]
    \caption{Proposed algorithm for solving problem $(\mathcal{P}_{A})$.}
    \label{alg1}
    \begin{algorithmic}[1]
    \STATE \textbf{Initialization}: Set initial $ \mathbf{A}^{(j)}, \mathbf{Q}^{(j)}$, $\mathbf{u}_k^{(j)}$, $\mathbf{v}^{(j)}$, $\boldsymbol{\Phi}^{(j)}$,  and iteration number $j=0$.
    \STATE \textbf{Step 1}: (Offline design with the S-CSI prior to the flight):
    \STATE \quad Update $\boldsymbol{\Phi}^{(j)}$ by solving $ (\mathcal{P}_{B1}) $;
    \STATE \quad Update $\mathbf{A}^{(j)}$ by solving $ (\mathcal{P}_{B2}) $;
    \STATE \quad Update $(\mathbf{Q}^{(j)}, \mathbf{u}_k^{(j)}, \mathbf{v}^{(j)})$ by solving $ (\mathcal{P}_{B3}) $;
    \STATE \quad Set $j\gets j+1$ and return to \textbf{Step 1}. Repeat the above until the convergence criteria is met.
    \STATE \textbf{Step 2}: (Online design with the effective I-CSI acquired at each time slot $n \in \mathcal{N}$):
    \STATE \quad Update $\mathbf{w}[n]$ by using \eqref{w};
    \STATE \quad Update $a_{k}[n]$ by solving $(\mathcal{P}_{C1})$.
    \end{algorithmic}
    \end{algorithm}

\section{Numerical Results}
In this section, numerical results are provided to evaluate the proposed hybrid design, and the simulation parameters are set as: $K=4$ and the horizontal coordinates of the users are $(-120, 10)$ $\rm{m}$, $(-80, 80)$ $\rm{m}$, $(80, 80)$ $\rm{m}$ and $(120, 10)~\rm{m}$, respectively. The altitude and horizontal coordinate of the RIS are $40$ $\rm{m}$ and  $(0, 0)$ $\rm{m}$, respectively. Besides, $M=M_x\times M_z =20\times20$, $N_t=5$, $I=100$, $\mathbf{q}_0 = [-500, 20]^T$ $\rm{m}$, $\mathbf{q}_F = [500, 20]^T$ $\rm{m}$,   $z_F=80~\rm{m}$,  $v_{max}=25$ $\rm{m/s}$, $\delta_t=1$ $\rm{s}$, $\sigma^2=-80$ $\rm{dBm}$, $d=\frac{\lambda}{2}$, $\alpha=2.2$, $\gamma=2.4$, $\kappa=3.5$, $\rho=-25$ $\rm{dB}$,  $P=0.01~\rm{W}$,  $\beta_{U\! G}=0$, and we set $\beta_{U\! R}=\beta_{R\! G}=\beta$ to facilitate the subsequent elaboration.

\begin{figure*}
	\setlength{\abovecaptionskip}{-5pt}
	\setlength{\belowcaptionskip}{-10pt}
	\centering
	\begin{minipage}[t]{0.33\linewidth}
		\centering
    \includegraphics[width=2.25in]{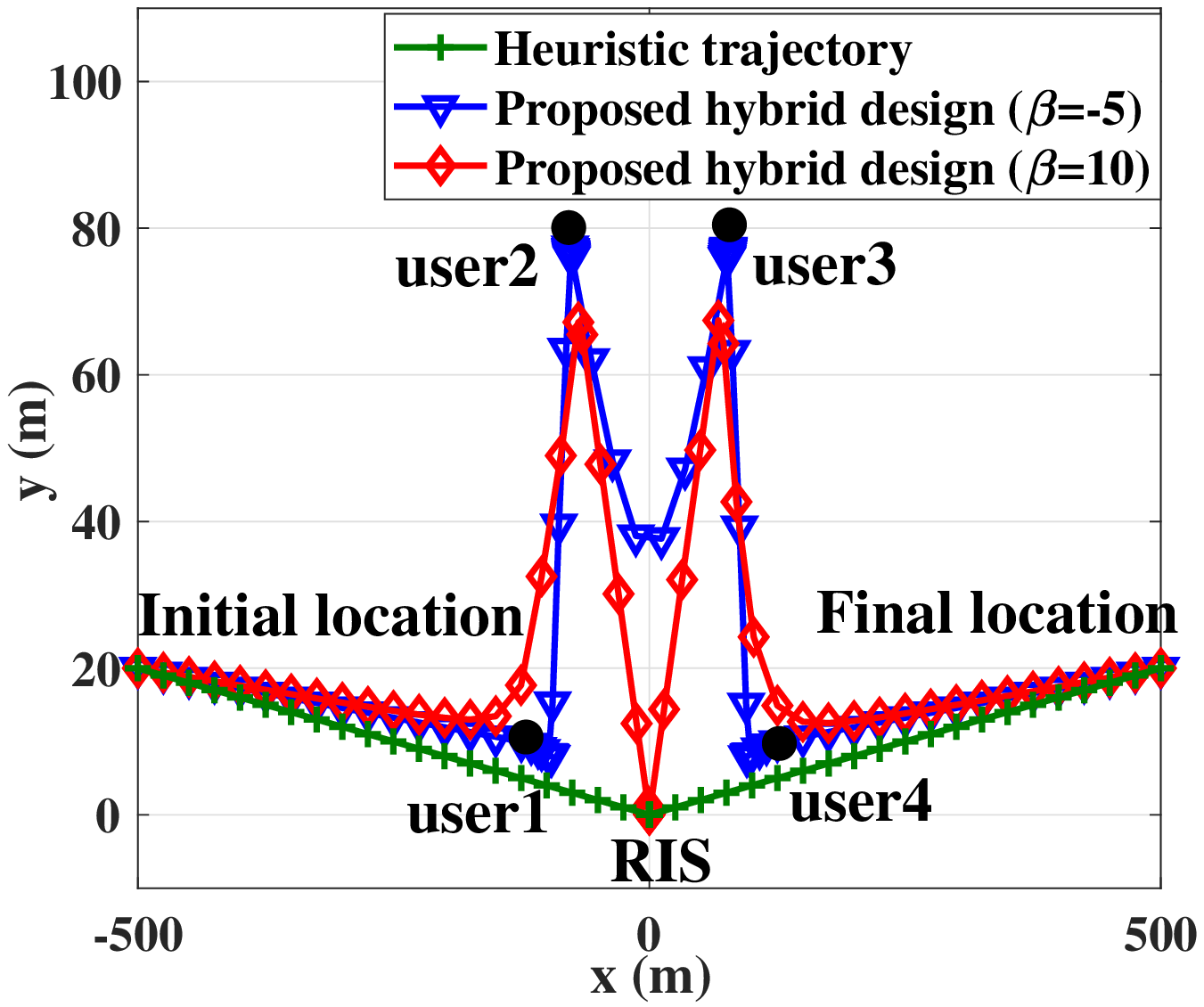}
		\caption{UAV trajectories.}
		\label{figure1}
	\end{minipage}%
	\begin{minipage}[t]{0.33\linewidth}
		\centering

      \includegraphics[width=2.2in]{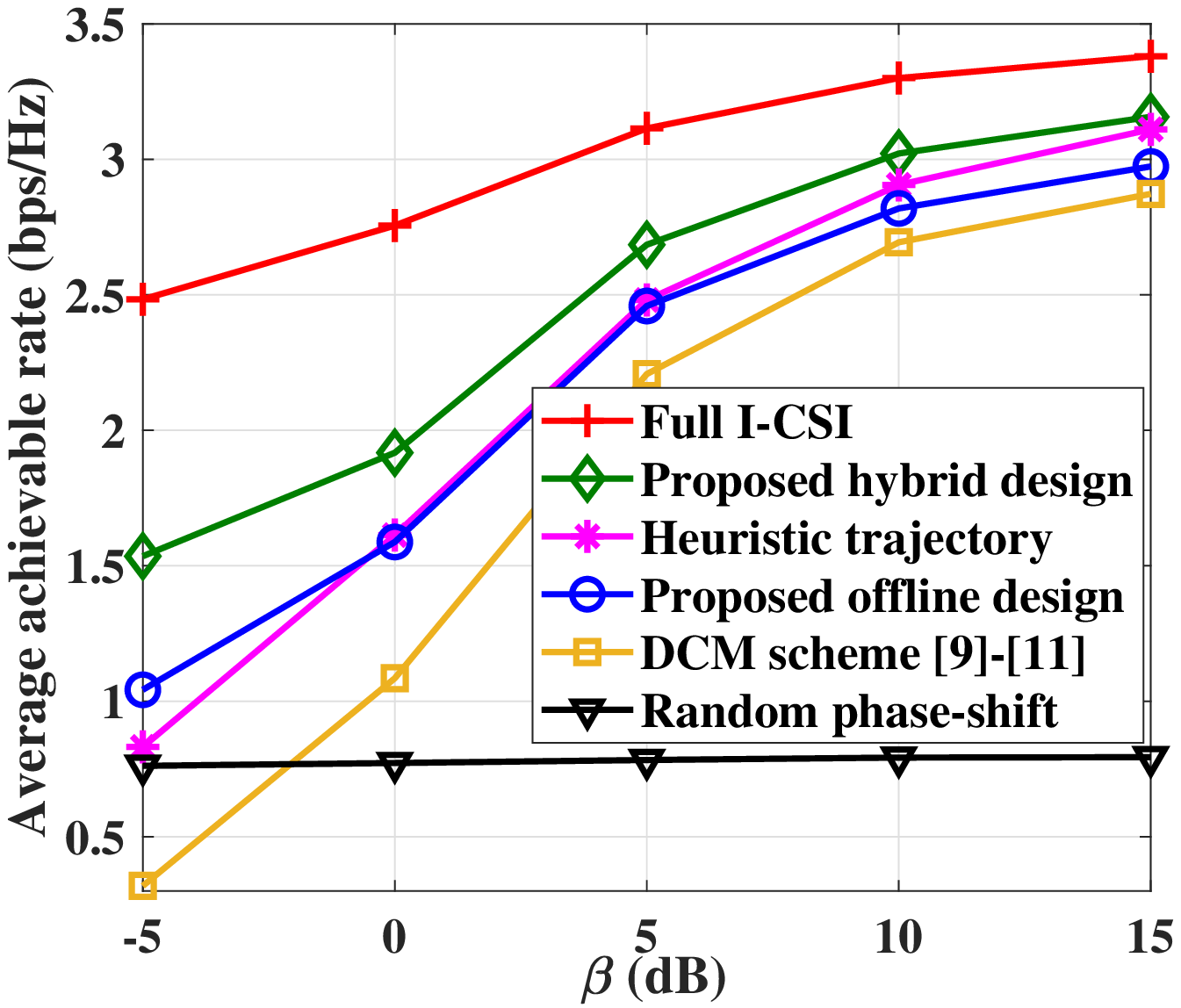}
		\caption{Achievable achievable rate versus $\beta$.}

		\label{figure2}
	\end{minipage}
	\begin{minipage}[t]{0.33\linewidth}
		\centering
   \includegraphics[width=2.2in]{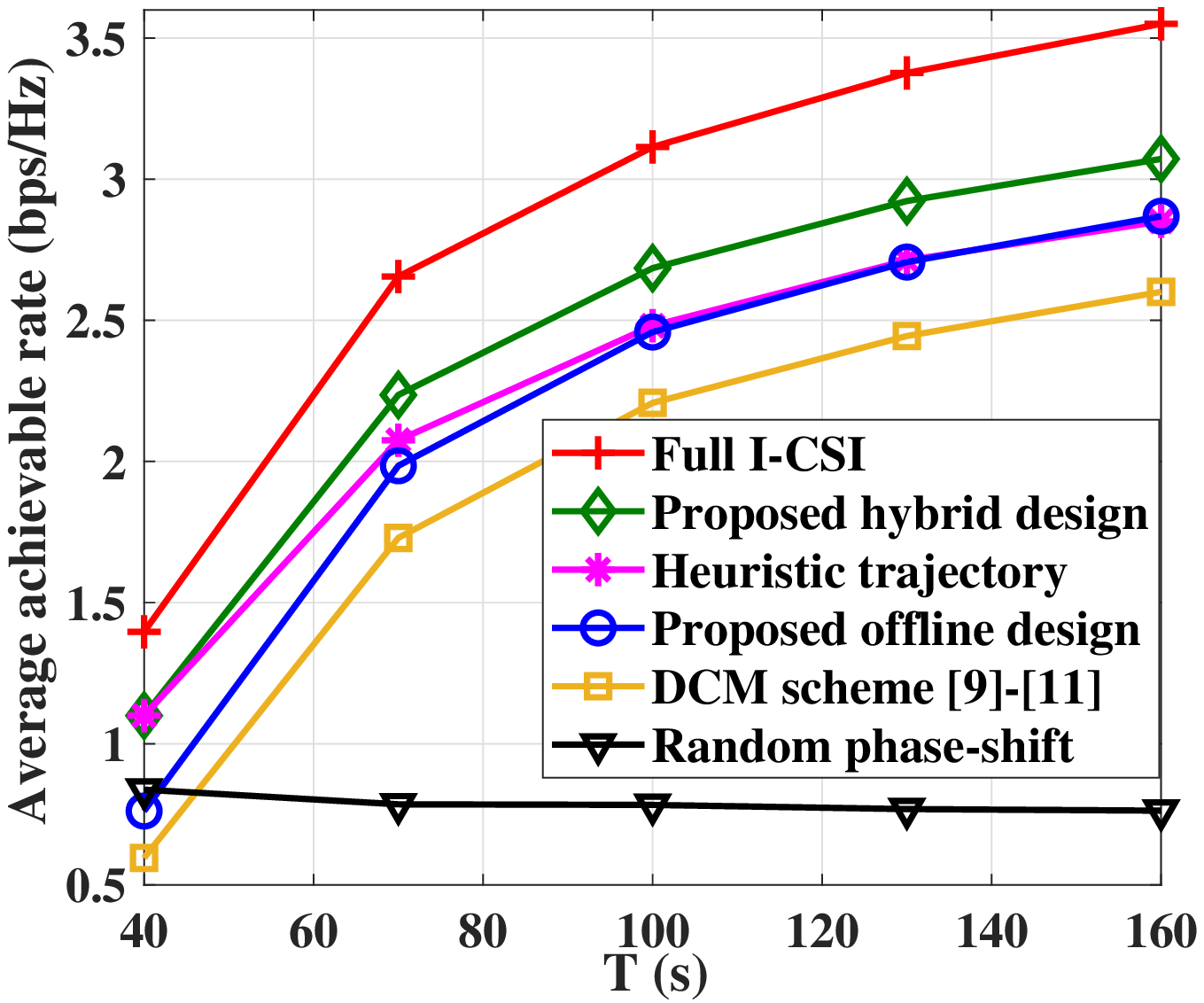}
		\caption{Achievable achievable rate versus $T$.}
		\label{figure3}
	\end{minipage}
\end{figure*}

      The following benchmark schemes are used for comparison. 1) Full I-SCI scheme: we consider an ideal scheme where the full I-CSI can be obtained over \emph{T} prior to the UAV's flight, which serves as the performance upper bound of $(\mathrm{P_A})$. 2) Proposed offline scheme: the proposed scheme without online design. 3) Deterministic channel model (DCM) scheme (methods in \cite{li2021ro, hua2021, Z2021su }): offline optimize the variables based on the deterministic Rician fading channel model. 4) Heuristic trajectory scheme: the UAV first flies to the RIS at $v_{\max}$, then hovers there as long as possible, and finally flies to the $\mathbf{q}_F$ at $v_{\max}$ for the rest of $T$. 5) Random  phase-shift scheme: proposed hybrid scheme with random RIS's phase-shift.

Fig. 1 shows the UAV trajectories by various algorithms with $T=100~\rm{s}$.  We  observe that, when the Rician factor of the proposed hybrid design is large ($\beta=10~\rm{dB}$), the UAV tries to approach each of the users in its route, but keeps static nearby the RIS as long as possible. This is because the UAV enjoys the gain provided by the RIS and users. However, when the Rician factor is small ($\beta=-5~\rm{dB}$), the UAV tries to approach each of all the four users and hovers above them instead of the RIS. This is because the signals from different transmission paths can not be aligned at users and the gain of the RIS deteriorates severely, thus the UAV flies to the users to obtain the channel gains of direct links.

In Fig. 2, we show the average achievable rate for different schemes versus $\beta$ with $T=100~\rm{s}$. It is observed that the average achievable rate increases with the Rician factor of all the schemes, because the related channels become more deterministic, and a larger proportion of the  LoS components can be obtained to improve the average achievable rate.  It is also observed that the hybrid design scheme outperforms the DCM scheme, especially when $\beta$ is small, since the hybrid design fully exploits the S-CSI and I-CSI. This indicates that the hybrid design is a better choice for this system, especially in the complex and random urban environments. Besides, the performance gap between the hybrid design scheme and heuristic trajectory scheme reduces as  $\beta$ increase, because the UAV prefers to be close to the RIS when $\beta$ is large.

In Fig. 3, we show the average achievable rate for different schemes versus $T$ with $\beta=5~\rm{dB}$. We observe that the achievable rate increases with $T$, since the UAV can fly to more favorable hovering locations to boost the system performance. Besides, the proposed hybrid design exceeds the heuristic trajectory and random phase-shift schemes. This shows that, by optimizing  the UAV trajectory and RIS's phase-shift, the system performance can be effectively improved.


    \section{Conclusions} \label{Conclusion}
In this letter, we investigated the RIS-aided UAV communication systems under the general Rician fading channel model and proposed the hybrid offline-online design to solve the maximal average achievable rate problem efficiently. Specifically, the offline design determines the UAV trajectory and RIS's phase-shift based on the S-CSI. The online design adjusts the user scheduling and transmit beamforming based on the effective I-CSI. Simulation results demonstrate  the effectiveness of the proposed hybrid design.

\end{document}